\documentclass[a4paper,floatfix,aip,jcp,reprint]{revtex4-1}
\usepackage{amsmath,amsthm,latexsym,amssymb,amsfonts,epsfig}
\usepackage[english]{babel}
\usepackage[utf8]{inputenc}			% Windows encoding
\usepackage{graphicx, texdraw}   %for figures
\usepackage{color}
\usepackage{bm}
\usepackage{ulem}  %STRIKE OUT
\usepackage{enumerate}
\usepackage{mathtools}% http://ctan.org/pkg/mathtools
\usepackage[colorlinks=true,citecolor=red, linkcolor=blue]{hyperref}
\newcommand{\pd}[2]{\frac{\partial #1}{\partial #2}} %partial derivative
\newcommand{\bc}{\mathbf{c}}

\newcommand{\bff}{\mathbf{f}}
\newcommand{\bg}{\mathbf{g}}

\newcommand{\br}{\mathbf{r}}

\newcommand{\bv}{\mathbf{v}}

\newcommand{\BR}{\mathbf{R}}
\newcommand{\BD}{\mathbf{D}}

\newcommand{\BF}{\mathbf{F}}
\newcommand{\BT}{\mathbf{T}}
\newcommand{\BE}{\mathbf{E}}
\newcommand{\BV}{\mathbf{V}}

\newcommand{\BZ}{\mathbf{Z}}
\newcommand{\BG}{\mathbf{G}}
\newcommand{\BA}{\mathbf{A}}
\newcommand{\BB}{\mathbf{B}}
\newcommand{\BC}{\mathbf{C}}

\newcommand{\BS}{\mathbf{S}}
\newcommand{\ex}{\hat{\mathbf{e}}_x}
\newcommand{\ey}{\hat{\mathbf{e}}_y}
\newcommand{\ez}{\hat{\mathbf{e}}_z}

\newcommand{\bzz}{\bm{\zeta}}
\newcommand{\ztt}{\bm{\zeta}^{tt}}

\newcommand{\zrr}{\bm{\zeta}^{rr}}

\newcommand{\mi}{\bm{\mu}}

\newcommand{\integr}[3]{\int_{#1}^{#2} \!\! \mathrm{d} {#3}\,}

\newcommand{\bee}{\begin{eqnarray}}
\newcommand{\eee}{\end{eqnarray}}
\newcommand{\oOmega}{\bm{\Omega}}

\newcommand{\tblu}{}

\newcommand{\n}{\hat{\mathbf{{n}}}}
\newcommand{\uu}{\hat{\mathbf{u}}}

\begin{document}

\title{Near-wall diffusion tensor of an axisymmetric colloidal particle}

\author{Maciej Lisicki}
\email{m.lisicki@damtp.cam.ac.uk}
\affiliation{Department of Applied Mathematics and Theoretical Physics, University of Cambridge, Cambridge, United Kingdom}
\affiliation{Institute of Theoretical Physics, Faculty of Physics, University of Warsaw, Warsaw, Poland}
\author{Bogdan Cichocki}
\affiliation{Institute of Theoretical Physics, Faculty of Physics, University of Warsaw, Warsaw, Poland}
\author{Eligiusz Wajnryb}
\affiliation{ Institute of Fundamental Technology Research, Polish Academy of Sciences, Warsaw, Poland}
\date{\today}

\begin{abstract}
Hydrodynamic interactions with confining boundaries often lead to drastic changes in the diffusive behaviour of microparticles in suspensions. For axially symmetric particles, earlier numerical studies have suggested a simple form of the near-wall diffusion matrix which depends on the distance and orientation of the particle with respect to the wall, which is usually calculated numerically. In this work, we derive explicit analytical formulae for the dominant correction to the bulk diffusion tensor of an axially symmetric colloidal particle due to the presence of a nearby no-slip wall. The relative correction scales as powers of inverse wall-particle distance and its angular structure is represented by simple polynomials in sines and cosines of the particle's inclination angle to the wall. We analyse the correction for translational and rotational motion, as well as the translation-rotation coupling. Our findings provide a simple approximation to the anisotropic diffusion tensor near a wall, which completes and corrects relations known from earlier numerical and theoretical findings.
\bigskip

\noindent{\bf Published in:} J. Chem. Phys {\bf 145}, 034904 (2016).

\noindent\doi{10.1063/1.4958727}
\end{abstract}

\maketitle

\section{Introduction}

Boundaries and interfaces are omnipresent in the colloidal world \cite{lang2015soft}. Geometric confinement introduces anisotropy in the diffusive motion of sub-micron particles, and the presence of neighbouring walls leads to a general slow-down of Brownian motion due to hydrodynamic interactions of the diffusing particle with boundaries \cite{HappelBrenner}. The central quantity in this context is the near-wall hydrodynamic mobility tensor $\mi$ which is related to the diffusion tensor $\BD$ by the fluctuation-dissipation theorem
 \begin{equation}
 \BD=k_B T \mi.
 \end{equation} 

{ Recent years have brought significant advancement in experimental techniques which allow to explore near-wall dynamics in more detail, including optical microscopy \cite{Banerjee2005,Kihm2004,Prieve1987,Prieve1999,Walz1995,Lin2000} and scattering techniques, such as evanescent wave dynamic light scattering \cite{Lan1986,Holmqvist2006,Holmqvist2007}. The latter is now a well-established tool which has profitably been used to investigate translational \cite{Lisicki2012} and rotational diffusion \cite{Rogers2012,Lisicki2014} of spherical colloids in dilute suspensions. Due to the complex nature of the experiments, available experimental data for non-spherical particles such as dumbbells \cite{Haghighi2013} or rods are still lacking proper interpretation. It is therefore particularly important in this context to understand the nature of hydrodynamic interactions of an axially symmetric particle with a wall, and has partially motivated this work.} Axisymmetric particles moving close to a boundary experience an additional anisotropic drag force on top of their own friction anisotropy stemming from their non-spherical shape. This coupling leads to a complicated behaviour, observed e.g. in simulations of such particles sedimenting next to a vertical wall \cite{Russel1977,Mitchell2015}, with the mobility of the particle depending on its position and orientation. \tblu{Available predictions for the near-wall mobility of an axisymmetric particle mostly feature a slender-body approach, yielding quite complex results for general wall-particle orientations near a wall \cite{Katz1975} or a fluid-fluid interface \cite{Yang1983},  analysed in detail in the context of sedimentation in several special alignments \cite{DeMestre1975}. On the other hand, previous numerical works involve the boundary integral method \cite{Hsu1989}, finite element method \cite{DeCorato2015} or stochastic rotation dynamics \cite{Padding2010} from which empirical relations are extracted.} The lack of theoretical predictions for the near-wall mobility of a rod-shaped and non-slender particle in an arbitrary configuration requires the use of more precise numerical methods.  A possible way is to use advanced algorithms involving bead-models which take into account lubrication when the particles come close to the interface\cite{Cichocki2000}, which are rather costly.

In order to fill this gap, in this work we derive a general form of the dominant correction to the bulk friction tensor due to the presence of a nearby no-slip wall from which the mobility tensor is calculated. This allows to verify and correct earlier predictions in terms of distance and orientation of the particle. {Importantly, the correction is valid for all axisymmetric particles, not just slender ones, provided that their bulk hydrodynamic properties are known.} Our analysis leads to a convenient representation of the mobility tensor in situations when the particle is moderately far from the wall.

 For the characteristic length of the body $L$, the relative correction scales as $(L/H)^\alpha$, where $H$ is the wall-particle distance, and the exponent $\alpha=1,2,3$ depends on the component of the friction matrix (translational, rotational, or coupling terms). We provide explicit analytical expressions for the dominant correction to bulk translational and rotational parts of the friction tensor which are the main result of the paper. We use them to calculate the corrections to the friction tensor of an axially symmetric particle explicitly in terms of $H$ and the particle's inclination angle $\theta$. By inverting the friction tensor, we then calculate the near-wall mobility tensor.

{The paper is organised as follows. First, we introduce the notion of friction and mobility tensors for a colloidal particle in Sec. \ref{sec1}. In Sec. \ref{sec:correction}, we sketch the idea behind the derivation of the correction, which is then given explicitly in Sec. \ref{sec:axisymm} for axially symmetric particles. The theoretical predictions are compared to numerical simulations for an exemplary case in Sec. \ref{simulation}, followed by conclusions in Sec. \ref{conclusions}. Appendix \ref{Multipole} contains the details of the multipole method and a description of the simulation method. Details of the derivation of the correction are given in Appendix \ref{Derivation}.}
\section{Near-wall friction and mobility tensors}\label{sec1}
 
We consider a single colloidal particle immersed in an incompressible Newtonian solvent of shear viscosity $\eta$. The configuration of the system is described by the position of the centre of the particle $\BR$ and its orientation which, for an axially symmetric particle, is specified by the unit vector $\uu$ pointing along the particle's symmetry axis. \tblu{On the colloidal length scales and for time scales typical e.g. for scattering experiments,  inertia of the fluid and the particle can be neglected.} The flow field $\bv(\br)$ around the particle is then described by the \tblu{stationary} Stokes equations\cite{KimKarrila}
\begin{equation}\label{stokes}
 \eta{\nabla}^2\bv(\br) -{\nabla} p(\br) = - \bff(\br),\qquad {\nabla}\cdot\bv(\br) = 0,
\end{equation}
where $\bff(\br)$ \tblu{is the force density the particle exerts on the fluid when subjected to flow}, and $p(\br)$ stands for modified pressure field which includes the effect of gravity. \tblu{The flow disturbances caused by the presence of the particle in confined geometry are affecting the motion of the particle itself.} These dynamic, solvent-mediated hydrodynamic interactions (HI) are long-ranged and have a pronounced effect on the dynamics of colloidal systems. This flow field may be superposed with an ambient linear flow $\bv_0(\br)$ satisfying the homogeneous Stokes equations, with the vorticity and rate of strain defined at a point $\br$ as \begin{equation}
\bm{\omega}_0(\br) = \tfrac{1}{2}\nabla\times\bv_0(\br), \qquad \BE_0(\br) = \overline{\nabla\bv_0(\br)},
\end{equation}
with the bar denoting the symmetric and traceless part.

\tblu{Given the force density, one can calculate the force, torque, and symmetric dipole moment (stresslet) exerted by the fluid on the particle according to
\begin{eqnarray}
\BF &=& -\integr{\Sigma}{}{\br}\bff(\br), \\
\BT &=& -\integr{\Sigma}{}{\br}(\br-\BR)\times\bff(\br), \\
\BS &=& -\integr{\Sigma}{}{\br}\overline{(\br-\BR)\bff(\br)},
\end{eqnarray}
where the integrals are performed over the particle surface $\Sigma$. Higher-order moments are defined in an analogous way.} In result of the external flow, motion is induced, and the particle gains linear and angular velocities, $\BV$ and $\bm{\Omega}$, respectively. Owing to linearity of the Stokes equations, the force moments $\BF$, $\BT$ and $\BS$, are linearly related to the velocity moments via the generalised friction (or resistance) tensor\cite{KimKarrila,EkielJezewska2009}
\begin{equation}\label{friction_single}
\begin{pmatrix}
\BF \\
\BT \\
\BS
\end{pmatrix} =  
\begin{pmatrix}
\bzz^{tt} & \bzz^{tr} & \bzz^{td} \\
\bzz^{rt} & \bzz^{rr} & \bzz^{rd} \\
\bzz^{dt} & \bzz^{dr} & \bzz^{dd} \\
\end{pmatrix}
\begin{pmatrix}
\bv_0(\BR)-\BV  \\
 \bm{\omega}_0(\BR)-\oOmega \\
\BE_0(\BR)
\end{pmatrix}.
\end{equation} 
Above we have decomposed the generalised friction tensor into 9 sub-matrices. The indices $tt$ and $rr$ denote the translational and rotational parts, respectively. The tensors $\bzz^{tr}$ and $\bzz^{rt}$ describe the translation-rotation coupling, and the tensors with superscript $d$ describe the response of the particle to an external elongational flow. In most cases, it is sufficient to consider only the $6\times 6$ friction matrix $\bzz$ relating the force and torque to linear and angular velocities. Here, we extend the friction matrix to the symmetric dipole moment subspace, since these elements turn out to be essential for the calculation of the correction to the friction matrix in the presence of a wall. 

In a complementary problem, if the forces and torques are known, the particle motion may be resolved by determining the mobility tensor ${\mi}$  which is related to the friction tensor by inversion
\begin{equation} \label{frictionmobility}
\mi = \begin{pmatrix}
\mi^{tt} & \mi^{tr} \\
\mi^{rt} & \mi^{rr} 
\end{pmatrix} = 
\begin{pmatrix}
\bzz^{tt} & \bzz^{tr} \\
\bzz^{rt} & \bzz^{rr} 
\end{pmatrix}^{-1}  = \bzz^{-1}.
\end{equation}
Using the Lorentz reciprocal theorem \cite{KimKarrila}, one may prove the symmetry properties of the mobility tensors. In a bulk system, the mobility tensor and the friction tensor, denoted by $\mi_0$ and $\bzz_0$, respectively, do not depend on the position of the particle due to translational invariance.

 The situation is different if a confining boundary is present, since symmetry is broken and the hydrodynamic tensors depend both on the distance to the boundary, and on the relative orientation of the particle with respect to the surface. The  friction tensors of a near-wall particle, $\bzz_w$, may be written as
 \begin{equation}\label{correctio}
  \bzz_w = \bzz_0 + \bm{\Delta}\bzz_w.
 \end{equation}
In the course of this work, we derive analytic formulae for the first-order approximation to $\bm{\Delta}\bzz_w$, with the expansion parameter being $L/H$, the ratio of the characteristic size of the particle, $L$, to the wall-particle distance $H$. By inverting $\bzz_w$ from Eq. \eqref{correctio}, we arrive at a convenient approximation to the near-wall mobility $\mi_w = \bzz_w^{-1}$.

\section{Method of the derivation}\label{sec:correction}

In order to determine the flow around a particle in a half-space bounded by an infinite, planar wall at $z=0$, one has to solve the Stokes equations (\ref{stokes}) with the no-slip boundary condition $\bv=0$ at the surface. \tblu{Due to linearity, Eq. \eqref{stokes} can be transformed into the integral form
\begin{equation}\label{boundintegral}
\bv(\br) = \bv_0(\br) + \integr{}{}{\br'} \BT(\br,\br')\cdot\bff(\br').
\end{equation}}
 For an unbounded fluid, the Green's function $\BT(\br,\br')=\BT(\br-\br')$ is the Oseen tensor \cite{KimKarrila} $
\BT_0(\br) = \left(\bm{1} + \hat{\br}\hat{\br}\right)/{8\pi\eta r}$, with $r=|\br|$ and $\hat{\br} = \br/r$. In the presence of boundaries, the full Green's tensor contains and additional part describing the flow reflected from interfaces. For a hard no-slip wall, the Green's tensor has been first found by Lorentz\cite{Lorentz1907} in 1907 as $\BT(\br,\br') = \BT_0(\br-\br') + \BT_w(\br,\br')$, where the wall contribution reads
\begin{align}\label{tensorblake}
\BT_w (\br,\br')=&  -\BT_0(\br-\br'^*)\cdot\mathbf{P} \\\nonumber
  &- 2 h \ez\cdot\BT_0(\br-\br'^*)\overleftarrow{\bm{\nabla}}_\br\cdot\mathbf{P},  \\\nonumber
   &+ h^2 {\nabla_\br^2}\BT_0(\br-\br'^*)\cdot\mathbf{P}
\end{align}
for a point force at a distance $h$ from a wall at $z=0$. \tblu{For a free surface, the wall-interaction part only contains the first term of the RHS of \eqref{tensorblake}}. Here, $[\mathbf{a}\overleftarrow{\bm{\nabla}}_\br]_{\alpha\beta} = \pd{}{r_\beta} {a}_\alpha $ and $\mathbf{P} = \bm{1} - \ez\ez$ denotes the reflection operator which transforms any point into its mirror image with respect to the wall. The asterisk denotes the mirror image, i.e. $\br'^\ast = \mathbf{P}\cdot\br'$. This expression has been interpreted by Blake\cite{Blake1971} in terms of the method of images for Stokes flows. The image system in this case involves three fundamental singularities: the reflection of the original Stokeslet and the so-called Stokeslet doublet and source doublet. However, since their amplitude is proportional to the wall-particle distance $h$, asymptotically they die out with distance $r$ as $1/r$. The tensor $\BT_w$ is often referred to as Blake's tensor. It can also be recast in different forms\cite{Cichocki2000}.

\tblu{The idea of the derivation relies on the expansion of the Blake's tensor \eqref{tensorblake} about the line connecting the centre of the particle and its hydrodynamic image. Thus the interaction between any two points of the particle may be represented by the vertical component of the Blake's tensor. By considering the action of higher force moments, higher-order flows incident on the particle can be found. Performing a multipole expansion of the resulting flow field\cite{Cichocki1998}, we project it onto the force multipole space and eventually find the explicit expressions for elements of the resistance matrix $\bzz_w$ which involve the elements of the bulk friction tensor $\bzz_0$ and the higher multipole elements of the Blake's tensor. The details of the derivation are presented in Appendices \ref{Multipole} and \ref{Derivation}. Below we present specific results for an axially symmetric particle.}

%===========================================================================
%
%				Section 4
%
%===========================================================================

%{\tred{\bf Comment on the nondiagonal components}.

%

\section{Dynamics of axisymmetric particles}\label{sec:axisymm}

It follows from symmetry properties that the bulk friction matrix of a general axisymmetric particle has a particular structure\cite{KimKarrila,Cichocki1988pch}. Moreover, if the particle has both axial and inversional symmetry $\uu\leftrightarrow-\uu$ (i.e. it is rod-like), its bulk friction matrix in Eq. \eqref{friction_single} simplifies, since $\bzz_0^{tr}$, $\bzz_0^{rt}$, $\bzz_0^{td}$ and $\bzz_0^{dt}$ vanish, i.e. translational motion is not coupled to torque or elongational flow in the bulk case. The correction has the following form
\begin{align}\label{corrtt} 
\bm{\Delta}\bzz^{tt}_w =& -\frac{1}{8\pi\eta}\frac{1}{2H} \mathbf{A}_1 
+ \frac{1}{(8\pi\eta)^2}\frac{1}{(2H)^2} \mathbf{A}_2 +\mathcal{O}(H^{-3}), \\
\bm{\Delta}\bzz^{tr}_w =& -\frac{1}{8\pi\eta}\frac{1}{(2H)^2} \mathbf{B} + \mathcal{O}(H^{-3}), \\
\bm{\Delta}\bzz^{rt}_w =& -\frac{1}{8\pi\eta}\frac{1}{(2H)^2} \mathbf{B}^\mathrm{T} + \mathcal{O}(H^{-3}), \\ \label{corrrr}
\bm{\Delta}\bzz^{rr}_w =& -\frac{1}{8\pi\eta}\frac{1}{(2H)^3}\mathbf{C} + \mathcal{O}(H^{-4}).
\end{align}
The form of the tensors $\mathbf{A}_{1,2}$, $\mathbf{B}$, $\mathbf{C}$ is derived from the multipole expansion of the vertical (axial) component of the Blake's tensor \eqref{tensorblake}  in Appendix \ref{Derivation}. As we have mentioned in Sec. \ref{sec1}, they depend on the $t,r,d$-components of the bulk generalised friction tensor of the particle and the particle's orientation via the inclination angle $\theta$. Explicitly, we write them in the following for an axisymmetric particle.

At contact, the elements of the friction matrix diverge which is expected physically, although no lubrication effects are included in this scheme. The mobility matrix is then obtained by inversion $\mi_w= \bzz_w^{-1} $, 
and by that we assure that the particle mobility decreases to a non-negative value at contact, but in a way different than with lubrication effects included.

We note that different elements of the friction matrix behave differently with distance. The hydrodynamic effect of the wall is most pronounced for translational motion where $\bm{\Delta}\bzz_w^{tt}\sim 1/H$. Rotational motion is least affected with $\bm{\Delta}\bzz_w^{rr}\sim 1/H^3$. This is in contrast with earlier numerical results due to Padding and Briels\cite{Padding2010}, \tblu{whose empirical finding was that both corrections scale as $1/H$, with a simple fitted angular dependence of the mobility components on $\theta$. Our findings provide exact expressions for the angular dependence, given as low-order polynomials in $\sin\theta$ and $\cos\theta$, correcting the previous empirical formulae. The source of the discrepancy lies in the inaccuracies of the numerical methods used in Ref. \onlinecite{Padding2010}.} They are also in complete agreement with the slender body expressions obtained by de Mestre {\it et al.}\cite{DeMestre1975} for the cases $\theta=\{0,\pi\}$ provided that the slender body results for $\bzz_0^{tt}$ and $\bzz_0^{rr}$ are used as input.

In order to discuss in detail the dependence of the correction terms on the orientation of the particle, we introduce two coordinate systems exploiting the symmetries of the problem, as sketched in Fig. \ref{rodimage}. The laboratory coordinate system (LAB) consists of three basis vectors $\left\{ \ex, \ey, \ez \right\}$, with the $z$-axis normal to the wall and the normal vector $\n=\ez$. The particle resides in the $xz$-plane. The rod-wall ({RW}) system is a body-fixed set of basis vectors $\left\{ \uu,\uu_{\perp 1},\uu_{\perp 2}\right\}$, where $\uu$ is the unit vector along the long axis of the particle, $\uu_{\perp 1}$ is parallel to the wall and perpendicular to the particle axis, and $\uu_{\perp 2}$ completes the orthonormal basis. The basis vectors are then given by $\uu_{\perp 1}=(\n\times \uu)/\left\vert \n\times \uu \right\vert$ and
$\uu_{\perp 2}=\uu_{\perp 1}\times \uu$. We note that $\cos \theta =\n\cdot\uu$.

For an axially symmetric particle, it is convenient to use the representation of the mobility matrix in the RW frame, in which the bulk tensors $\bzz^{tt}$ and $\bzz^{rr}$ are diagonal. The structure of the near-wall tensors is identical to that given in Ref. \onlinecite{DeCorato2015}. In the body-fixed frame of reference RW, the correction tensors in Eqs \eqref{corrtt}-\eqref{corrrr} may be explicitly written in terms of the inclination angle $\theta$. In the formulae, we find the elements of the bulk friction tensor of the particle, namely the coefficients of translational and rotational friction in the directions parallel and perpendicular to the body axis, given respectively by $\zeta^{\alpha}_\parallel =  \uu\uu:\bzz_0^{\alpha\alpha}$ and $\zeta^{\alpha}_\perp = (\bm{1}-\uu\uu):\bzz_0^{\alpha\alpha}$, where $\alpha=\{t,r\}$ and : denotes double contraction. In addition, the correction terms for rotational motion and rotation-translation coupling contain the coefficient $\zeta^{dr}$. As seen from Eq. \eqref{friction_single}, it quantifies the stresslet exerted on the particle rotating with a prescribed angular velocity. These coefficients can be taken from bulk results for friction of axisymmetric particles. For ellipsoidal particles, analytical expressions are available\cite{KimKarrila}, whereas for more complex shapes the coefficient may be determined using bead-\cite{Cichocki1994g} or shell-models\cite{Ortega2003}.

Now we can write the tensors in Eqs \eqref{corrtt}--\eqref{corrrr} explicitly in the body-fixed frame RW.  For the translational part \eqref{corrtt}, we find the correction's angular dependence as
\begin{align}\label{angulartt}  
&\mathbf{A}_1 =-\frac{3}{2}
\begin{pmatrix}
(\zeta^{t}_\parallel)^2(1+\cos^2\theta) & 0 & -\zeta^{t}_\parallel\zeta^{t}_\perp \sin\theta\cos\theta \\
0 & (\zeta^{t}_\perp)^2 & 0 \\
-\zeta^{t}_\parallel\zeta^{t}_\perp \sin\theta\cos\theta & 0 & (\zeta^{t}_\perp)^2 (1 + \sin^2\theta) 
\end{pmatrix}, \\ 
&\mathbf{A}_2 = \frac{9}{4} 
\begin{pmatrix}
A_\parallel & 0 & A_{\parallel\perp} \\
0 &  A_{\perp 1} & 0 \\
A_{\parallel\perp} & 0 & A_{\perp 2}
\end{pmatrix}.
\end{align}
with $A_{\perp 1}=(\zeta^{t}_\perp)^3$ and
\begin{align} 
A_{\parallel} &=(\zeta^{t}_\parallel)^3(1+\cos^2\theta)^2 + \zeta^{t}_\perp(\zeta^{t}_\parallel)^2 \sin^2\theta\cos^2\theta, \\ \nonumber
A_{\perp 2} &= (\zeta^{t}_\perp)^3(1+\sin^2\theta)^2 + (\zeta^{t}_\perp)^2\zeta^{t}_\parallel \sin^2\theta
\cos^2\theta, \\ \nonumber
A_{\parallel\perp} &=-\zeta^{t}_\parallel\zeta^{t}_\perp[\zeta^{t}_\parallel(1+\cos^2\theta)+\zeta^{t}_\perp(1+\sin^2\theta)] \sin \theta\cos\theta.
\end{align}

We note here that since the axis perpendicular to the rod and parallel to the wall (in the direction of  $\uu_{1\perp}$) is invariant with respect to the LAB to RW frame transformation, the middle element of the matrix above is angle-independent. The translation-rotation coupling part reads
\begin{align}
\mathbf{B} =\frac{3\zeta^{dr}}{2}
\begin{pmatrix}
0 & \zeta^{t}_\parallel (1+\cos^2\theta) \sin\theta & 0 \\
0 & 0 & \zeta^{t}_\perp \cos\theta \\
0 & -\zeta^{t}_\perp (1+\sin^2\theta)\cos\theta & 0 
\end{pmatrix}.
\end{align}
Finally,  the rotational tensor $\mathbf{C}$ in Eq. \eqref{corrrr} is a sum of three contributions and has the form
\begin{align} \label{angularrr}
&\mathbf{C} = \\ \nonumber
&-\frac{1}{2}\begin{pmatrix}
(\zeta^{r}_\parallel)^2(5-3\cos^2\theta) & 0 & 3 \zeta^{r}_\parallel\zeta^{r}_\perp \sin\theta\cos\theta \\
0 & 5(\zeta^{r}_\perp)^2 & 0 \\
3 \zeta^{r}_\parallel\zeta^{r}_\perp \sin\theta\cos\theta  & 0 & (\zeta^{r}_\perp)^2(5 - 3 \sin^2\theta)
\end{pmatrix} + \\  \nonumber 
&\frac{3\zeta^{dr}}{2}\begin{pmatrix}
0 & 0 & \zeta^{r}_\parallel \sin\theta\cos\theta \\
0 & -2\zeta^{r}_\perp(1-2 \cos^2\theta) & 0 \\
\zeta^{r}_\parallel \sin\theta\cos\theta  & 0 & 2 \zeta^{r}_\perp \cos^2\theta
\end{pmatrix} + \\ \nonumber
&-\frac{3(\zeta^{dr})^2}{2}\begin{pmatrix}
0 & 0 & 0 \\
0 & 3+\cos^2\theta-\cos^4\theta & 0 \\
0  & 0 & 1+2\cos^2\theta
\end{pmatrix}.
\end{align}
For completeness of the discussion, it is worth noting that it is possible to find analytically the leading order behaviour of the mobility functions by expanding the inverted friction matrix \eqref{corrtt}-\eqref{corrrr}. In this way, the dominant terms of the correction may be evaluated as functions of $\theta$ and indicate a very simple angular relations for the components, namely low-order polynomials in $\sin\theta$ and $\cos\theta$. However, as we noted before, this is not an optimal strategy, since the mobility functions obtained by inversion may become negative when the particle approaches the wall ($H$ is small compared to $L$). Therefore, it proves better to first calculate the wall-corrected friction tensor, and then invert it to obtain the mobility tensor. 

\section{{Numerical results}}\label{simulation}

{In order to assess the applicability range of the correction, we compare our theoretical result to precise numerical simulations using the {\sc Hydromultipole} package\cite{Cichocki2000}. As a representative test example, we consider a rod-like particle of aspect ratio $p=L/D =10$ constructed out of $N=10$ spherical beads glued together along a straight line. For the needs of demonstration, we choose one inclination angle, $w=\cos\theta=0.5$, implying the minimal contact distance $H=0.275L$. We plot the components of the mobility matrix calculated using the procedure outlined above, and compare them to the corresponding accurate numerical predictions. To this end, we introduce the following notation in the body-fixed RE frame. }

 Taking into account the invariant properties of the rod-wall system and the Lorentz symmetry, we can write the translational part as
\begin{equation}
\label{mutt}
\mi_w^{tt}(z,\uu;\n)=%
\begin{pmatrix}
a_{t}(z,w) & 0 & c_{t}(z,w) \\ 
0 & b_{t}(z,w) & 0 \\ 
c_{t}(z,w) & 0 & d_{t}(z,w)%
\end{pmatrix}%
_\mathrm{RW}
\end{equation}
where $w=\cos \theta $. The matrix $\mi^{rr}(z,\uu;\n)$ has a similar structure with elements $a_{r}$, $b_{r}$, $c_{r}$, and $d_{r}$, respectively. In
both cases, the elements $a$, $b$ and $d$ are even functions of $w$ and the elements $c$ are odd ones. For the $tr$ part, we have
\begin{equation} \label{mutr}
\mi_w^{tr}(z,\uu;\n)=%
\begin{pmatrix}
0 & a_{tr}(z,w) & 0 \\ 
b_{tr}(z,w) & 0 & c_{tr}(z,w) \\ 
0 & d_{tr}(z,w) & 0%
\end{pmatrix}%
_\mathrm{RW}
\end{equation}
The elements $a_{tr}$ and $b_{tr}$ are even functions of $w$, while $c_{tr}$ and $d_{tr}$ are odd ones. By taking the transposition of the above matrix we get the $(rt)$ part. 

\begin{figure}[h]
\centering
\includegraphics[width=0.5\textwidth]{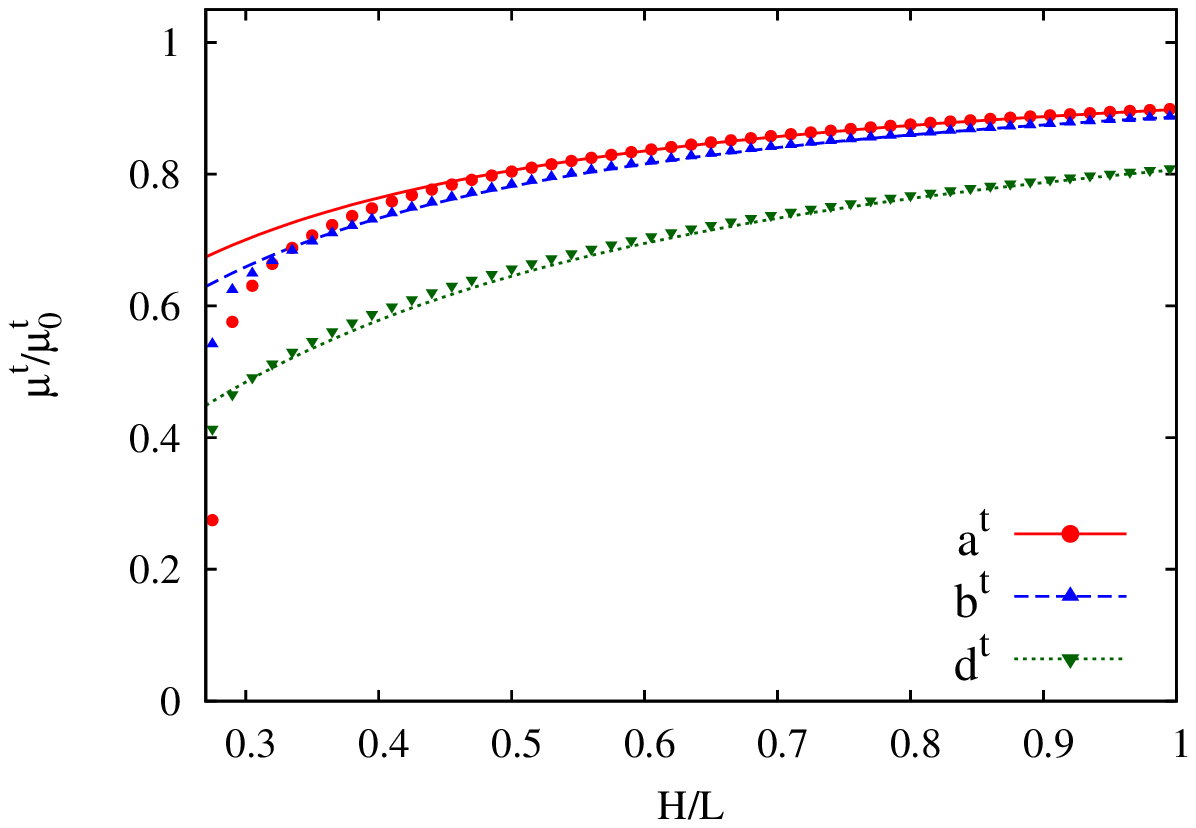}
\includegraphics[width=0.5\textwidth]{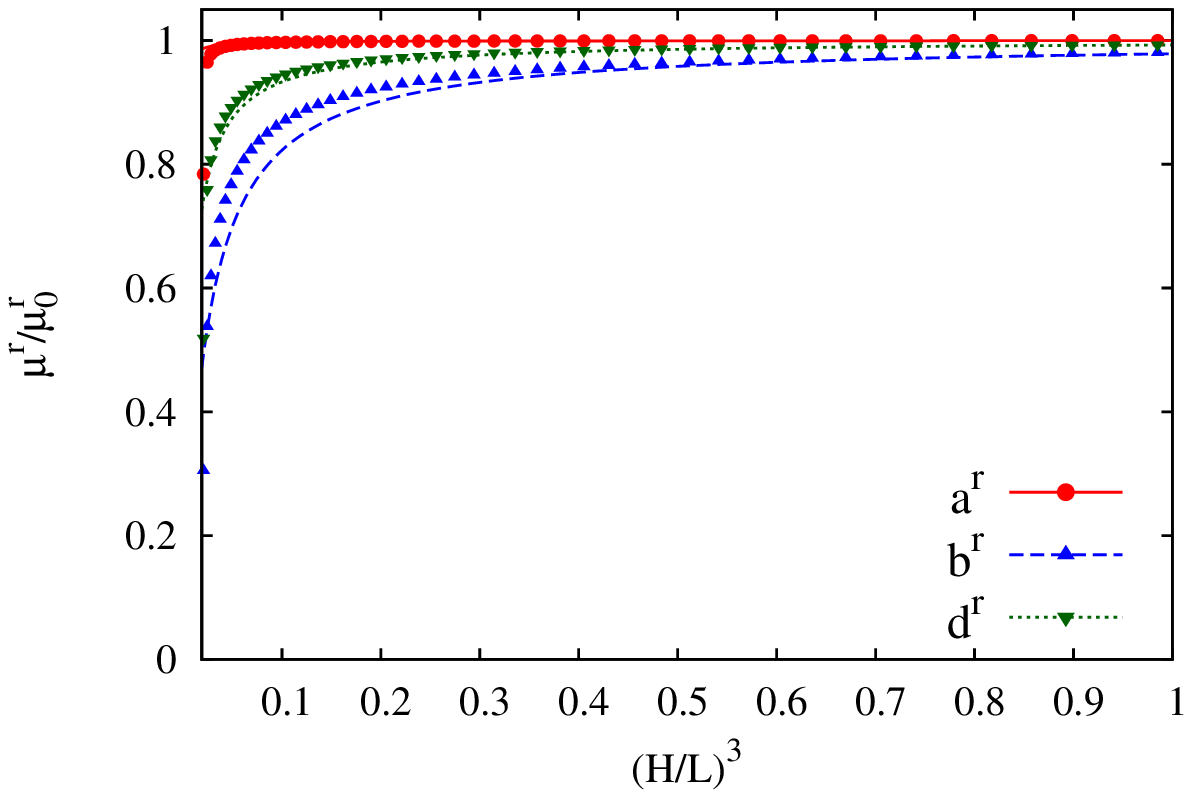}
  \caption{\label{comparison} \tblu{Comparison of the near-wall mobility of a rod of apect ratio $p=10$ at an angle $\cos\theta=0.5$ to the wall, as predicted by the correction (solid lines) and precise {\sc Hydromultipole} numerical simulations (data points). The coefficients are normalised by their corresponding bulk values, to that the all tend to unity at $H\to\infty$. Top: diagonal elements of the translational mobility matrix in the RW frame. Bottom: rotational diagonal elements. }}
\end{figure}

{The diagonal components of the translational and rotational diffusion tensor of the rod are plotted in Fig. \ref{comparison}. They reveal that for translational motion in this particular case the correction accurately represents the actual mobility even up to $H/L \approx 0.4$, and the asymptotic inverse-distance behaviour of the correction is evident. Similarly, the rotational components are even less sensitive to the effect of the wall due to the rapid decay of the HI for rotational motion. For larger distances, the mobility matrix obeys the necessary symmetries, with $b^t = d^t$ and $b^r = d^r$ asymptotically.}

\begin{figure}[h]
\centering
\includegraphics[width=0.5\textwidth]{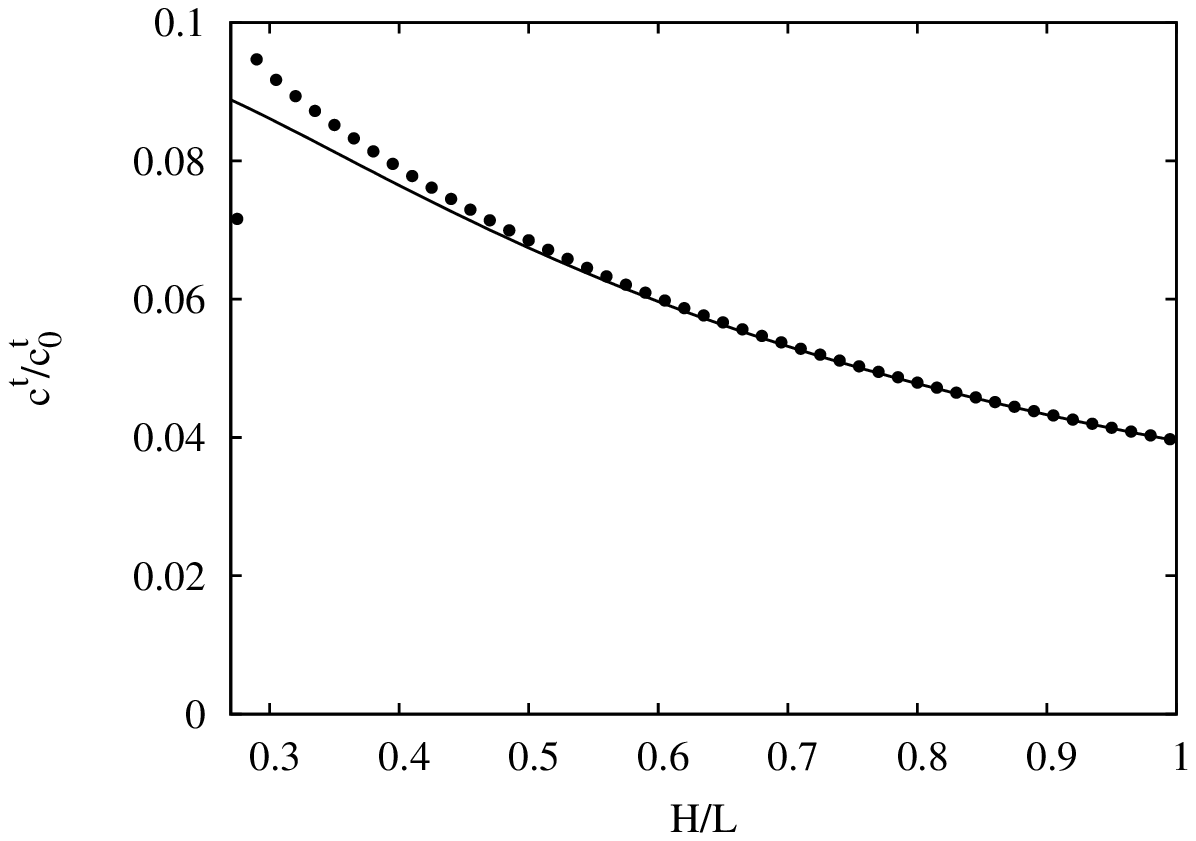}
\includegraphics[width=0.5\textwidth]{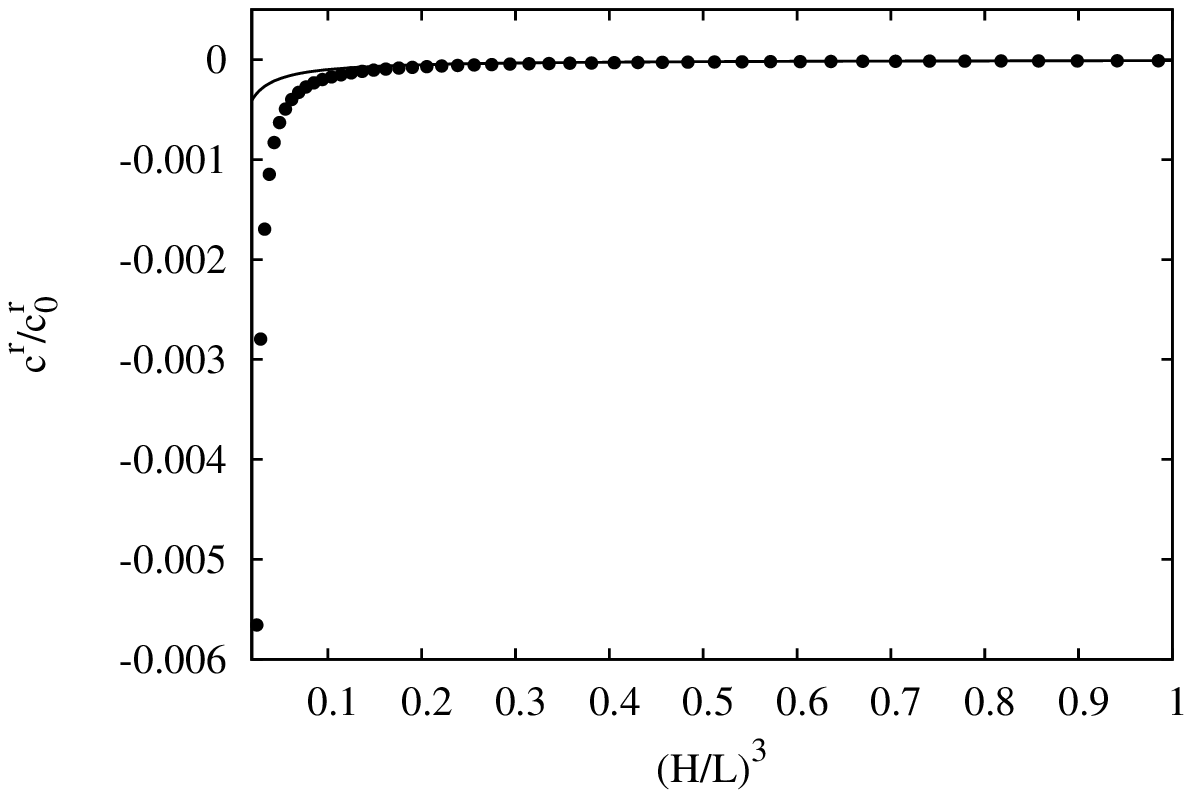}
  \caption{\label{comparison_nondiag}\tblu{ Non-diagonal components of the near-wall translational and rotational mobility tensors for the inclination angle $\cos\theta=0.5$. The data points are predictions of multipole simulations with lubrication included, while the solid lines are predicted by our analytical formulae for the correction. The coefficients are normalised by bulk average values of the diagonal terms, e.g. $c^{t,r}_0=\sqrt{\mu^{t,r}_\parallel \mu^{t,r}_\perp}$. }}
\end{figure}

{The presence of the wall introduces also non-diagonal component to the mobility tensors, which we depict in Fig. \ref{comparison_nondiag}. Normalised by the appropriate combinations of bulk mobility coefficients, these elements are rather small. Nevertheless, they numerical results are again in agreement with theoretical predictions up to quite close wall-particle distances, both for translations, and rotations.}

\begin{figure}[h]
\centering
\includegraphics[width=0.5\textwidth]{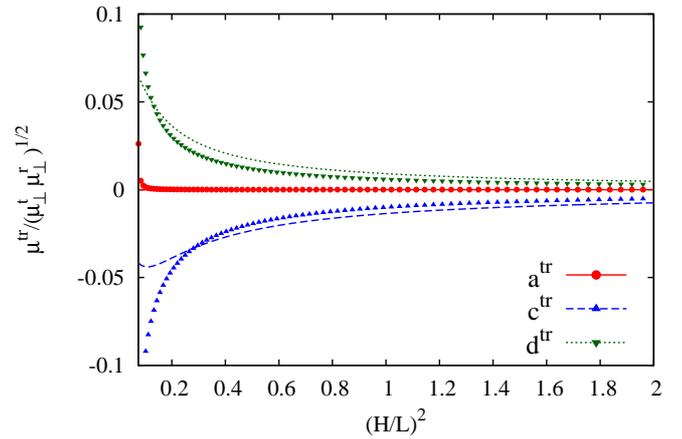}
  \caption{\label{comparison_tr} \tblu{Components of the  translation-rotation coupling mobility tensor $\mi^{tr}$ for the inclination angle $\cos\theta=0.5$. The data points are predictions of multipole simulations with lubrication included, while the solid lines are predicted by our analytical formulae for the correction. The coefficients are normalised by a combination of  bulk coefficients $\sqrt{\mu^{t}_\perp \mu^{r}_\perp}$. The deviations are most pronounced in the coupling tensor. The overall values are, however, rather small, not exceeding 5\% for $H=0.4L$.}}
\end{figure}

{The translation-rotation coupling tensors become more significant as the particle approaches the wall, as seen from Fig. \ref{comparison_tr}. Compared to the characteristic bulk quantities, however, they seem to play a marginal role in this case. With the derived correction, we are able to reproduce them accurately again up to $H/L\approx 0.5$.}

{The comparison of the correction to numerical results in the case of a relatively long ($p=10$) rod-like particle is quite favourable. Indeed, we expect the correction to work even better for more slender particles, since it can be shown analytically using the slender body results for the bulk friction that the relative correction to the translational mobility tensor (e.g. $a^t/a^t_0$ etc.) decreases slowly with increasing aspect ratio as $1/\log p$. }

\section{Conclusions}\label{conclusions}

We have presented a simple analytical scheme which allows for the representation of the near-wall friction and mobility tensors of a rod-like colloid close to a planar no-slip wall. \tblu{The correction to bulk mobility, expressed in terms of the bulk hydrodynamic properties of the particle, is valid for general axially symmetric colloids, which need not be slender.} Our results show that the distance dependence varies between the types of motion in focus (translational, rotational, and $tr$-coupling). Moreover, we have demonstrated by analytical formulae that near-wall friction and mobility for particles at moderate distances from the wall indeed depends on their orientation via simple polynomials in sine and cosines of the inclination angle $\theta$, as conjectured by Padding {\it et al.}\cite{Padding2010}.  By that we have also verified earlier theoretical developments and recent numerical predictions\cite{DeCorato2015}. \tblu{Our results are in agreement with numerical calculations even in the case when $L/H \sim O(1)$, rendering the results practical for large and moderate wall-particle distances.}

\begin{acknowledgments}

ML acknowledges support from the National Center of Science grant no. 2012/07/N/ST3/03120. Part of the research has been conducted under a David Crighton Fellowship awarded to ML at the University of Cambridge, and within the Mobility Plus Fellowship awarded to ML by the Polish Ministry of Science and Higher Education.
\end{acknowledgments}

\appendix

\section{\tblu{The multipole expansion}}\label{Multipole}

The idea of the multipole method relies on expressing the force densities and velocities on the surfaces of many spheres immersed in the fluid in the form of a boundary integral equation, which is then projected onto a complete set of multipolar solutions of the Stokes equations. The resulting system of linear equations may then be truncated and solved numerically for a conglomerate of spheres moving together. By projecting the many-particle friction matrix obtained in this way onto the subspace of rigid body motions of the conglomerate, the friction tensor of a complex-shaped particle is found. The method has been greatly developed over the last decades, and is presented in more details, e.g. in Refs. \onlinecite{Cichocki1994g,EkielJezewska2009}.

With the use of the concept of induced forces due to Bedeaux and Mazur \cite{Bedeaux1974}, the validity of the Stokes equations (\ref{stokes}) may be formally extended inside the particles by taking an appropriate surface distribution of the forces $\bff_i(\br)$ on the surfaces of the particles $i=1,\ldots,N$. For the stick boundary conditions, the velocities on the surfaces read
\begin{equation}
\bv_i(\br) = \BV_i + \oOmega_i\times(\br-\BR_i),
\end{equation}
and Eq. \eqref{boundintegral} on the surfaces of the particles takes the form
\begin{equation}\label{3_boundint1}
\bv_i(\br) = \bv_0(\br) + \sum_{j=1}^{N}\integr{}{}{\br'} \BT(\br,\br')\cdot\bff_j(\br'), \quad \br\in \Sigma_i
\end{equation}
where $\bv_0(\br)$ represents an ambient flow in the absence of the spheres.

We now separate the second term on the RHS of \eqref{3_boundint1} into the contribution from distinct particles and the self-contribution. The self part is found by considering a single particle $i$ in an ambient flow $\bv^\mathrm{in}$. The force density $\bff_i$ it exerts on the fluid is linearly related to the relative velocity at the surface, viz. $\bff_i = - \BZ_0(i)(\bv_i-\bv^\mathrm{in})$ where the integral operator $\BZ_0(i)$ is called the single-particle resistance operator, or the friction kernel \cite{JonesSchmitz1988}, and depends solely on the internal composition and surface properties of the particle \cite{Cichocki1988pch,JonesSchmitz1988}. For the distinct part ($i\neq j$), we introduce the Green's integral operator (propagator) $\BG(ij)$:
\begin{align} \label{propagator}
[\BG(ij)\bff_j](\br) & \equiv \integr{}{}{\br'} \BT(\br,\br')\cdot \bff_j(\br'), \qquad \br\in\Sigma_i,
\end{align}
which allows Eq. \eqref{3_boundint1} to be written as
\begin{equation}\label{3_boundint}
\bv_i-\bv_0 = \BZ_0^{-1}(i)\bff_i + \sum_{j\neq i}^N \BG(ij)\bff_j, \quad i=1,\ldots,N.
\end{equation}

The above equations can be transformed into an infinite set of algebraic equations by expanding the velocity field and induced force densities in a basic set of irreducible multipoles developed by Felderhof and co-workers \cite{Schmitz1978,Cichocki1988pch,Schmitz1982b,Schmitz1982d,Cichocki1994g}. For the velocity, the irreducible multipoles are linear combinations of Lamb's solution of the homogeneous Stokes equations\cite{KimKarrila}. They are labelled by three numbers: $l = 1,2,\ldots$, $m=-l,\ldots,l$ and $\sigma\in\{0,1,2\}$. The force multipoles can be likewise be described by the labels $(lm\sigma)$.  The details of the expansion, along with explicit form are given in Refs. \onlinecite{Cichocki1988pch,Cichocki2000,EkielJezewska2011}. 

We include the expansion coefficients of the velocities $\bv_i-\bv_0$ in an infinite-dimensional vector $\mathbf{c}$ which encompasses all the velocity multipoles for all the particles.  In a similar manner, we arrange the force multipole moments in the vector $\bff$. After the multipole expansion is performed, the integral operators $\BZ_0$ and $\BG$ become matrices and Eq. \eqref{3_boundint} is transformed into an algebraic equation
\begin{equation}\label{multipoleeqset}
\bc = (\BZ_0^{-1} + \BG)\cdot \bff.
\end{equation}
The multipole matrix elements of $\BZ_0$ for different particle models are given in Ref. \onlinecite{Cichocki1988pch}, while the elements of $\BG$ have been calculated in Ref. \onlinecite{Cichocki1998} for the case of an unbounded fluid, a fluid bounded by a free surface and a fluid bounded by a hard wall.

In the friction problem, the forces acting on the particles are sought, given their velocities. Upon inverting the above relation, the grand resistance matrix $\BZ$ is found as
\begin{equation}\label{3_frictionproblem}
\bff = \BZ \cdot \bc, \qquad \BZ = (\BZ_0^{-1} + \BG)^{-1}.
\end{equation}
The friction matrix defined in Eq. \eqref{friction_single} can  be found by projecting $\BZ$ on the subspaces $t,r,d$, corresponding to $(lm\sigma)$ equal to $(1m0)$, $(1m1)$, and $(2m0)$, respectively. For example, the force multipole $t$ with $l=1$ and $\sigma=0$ has three spherical components $m=-1,0,1$ corresponding to three components of the total force.

In numerical computations, infinite matrices $\BZ_0$ and $\BG$ in Eq. (\ref{multipoleeqset}) are truncated at the multipole order $\ell$, so that only the elements with $l\leq\ell$ are considered \cite{Cichocki1994g}. After such a truncation, the matrix $(\BZ_0^{-1} + \BG)$ is inverted, and the force multipoles are determined. To improve numerical convergence of this scheme, the obtained grand friction matrix $\BZ$ in Eq. (\ref{3_frictionproblem}) is additionally corrected for lubrication effects \cite{EkielJezewska2011,Cichocki1999lub,Durlofsky1987,Brady1988}. The matrix $\BZ$ constructed in the multipole method is not pairwise additive, and accounts fully for many-body hydrodynamic interactions. The approximation is introduced at the level of truncation of the multipoles, and its error may be controlled. 

The procedures outlined above have been implemented in a Fortran code {\sc Hydromultipole}  \cite{Cichocki1994g,Cichocki1999lub} by Wajnryb and collaborators. The method for calculating the near-wall hydrodynamic tensors has been laid out by Cichocki {\it et al.} in Ref. \onlinecite{Cichocki2000}. We employ these codes to calculate the friction tensors of non-spherical particles represented by their bead-models. Once the friction matrix $\bzz$ is known, the mobility matrix $\mi$ is found by inversion.

\section{\tblu{Details of the derivation}}\label{Derivation}

For a wall-bounded fluid, it follows from the form of Eq. \eqref{tensorblake} that the propagator $\BG$ can be decomposed as
\begin{equation}\label{Gdecompose}
\BG= \BG_0 + \BG_w, 
\end{equation}
where the part $\BG_0$ consists of the multipole elements of the Oseen tensor\cite{Cichocki2000}, while $\BG_w$ describes the wall contribution, the multipole matrix elements of which are calculated for a free surface and a hard wall in Ref. \onlinecite{Cichocki1998} (see also Ref. \onlinecite{Cichocki2000}). 

In order to find the asymptotic correction to the bulk friction of a particle moving close to a wall, we employ the scattering expansion \cite{Felderhof1988}. We start from rewriting the grand resistance matrix in Eq. (\ref{3_frictionproblem}) in the following form
\begin{equation}\label{3_frictionproblem1}
\BZ = (\BZ_0^{-1}+\BG_0+\BG_w)^{-1}.
\end{equation}
%.
 When the wall-particle distance is considerably larger than the particle itself, so we expect the wall contribution $\BG_w$ to be a small correction. Expanding Eq. \eqref{3_frictionproblem1} yields the form of the correction to the bulk resistance matrix 
\begin{equation}\label{A_delta}
\bm{\Delta} = \BZ-\BZ_b =  -\BZ_b {\BG_w}\BZ_b + \BZ_b{\BG_w}\BZ_b{\BG_w}\BZ_b - \ldots
\end{equation}
with $\BZ_b$ being short for the bulk resistance matrix of the particle $(\BZ_0^{-1}+\BG_0)^{-1}$.  Further on, we evaluate the dominant terms of the correction for all the elements of the friction matrix in the $t,r$ subspace.

\begin{figure}
\centering
\includegraphics[width=0.75\columnwidth]{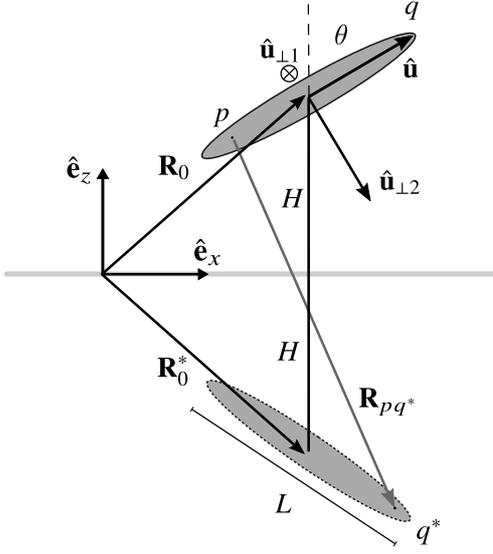}
\caption{\label{rodimage} The relevant coordinate systems and a schematic illustration of the expansion of the distance between interacting points of the particle (e.g. a rod) and its image. We depict the interaction between the point $p$ with the image $q^*$ of point $q$. For large wall-particle distances, it may be expanded around the vertical line connecting the particles' centres lying at a distance $2H$ apart, so that $\BR_{pq^*} \approx \BR_0-\BR_0^* = 2H\mathbf{e}_z$.}
\end{figure}

The propagator ${\BG_w}$ connects the beads building up the particles with the beads of the image particle. Consider two interacting beads $p$ and $q^*$ building up the conglomerate and its image, respectively, as illustrated in Fig. \ref{rodimage}. Introducing coordinates relative to the centre of each conglomerate, we have $\BR_p = \BR_0+\br_p$ and $\BR^*_{q} = \BR_0^* + \br^*_{q}$. Hence the distance between the particles may be written as
\begin{equation}\label{Rpq}
\BR_{pq^*} = \BR_p - \BR^*_q = 2H\n + (\br_p - \br_{q}^*),
\end{equation}
where we have used the fact that $\BR_0-\BR_0^* = 2H \n$. For the wall-particle distance $H$ large compared to the particle size $L$, and thus for $|\br_p-\br_q^*|\ll H$, we may expand the distance between each pair around the direction normal to the wall. Then, in leading order, the propagator takes the form ${\BG}_w(\BR=2H\n)$. Due to this fact, its multipole elements have the axial symmetry around the normal direction $\n$.  In this case, the bulk grand resistance matrix $\BZ_b$ reduces to the single-particle bulk friction matrix $\bzz_0$ as in Eq. (\ref{friction_single}). 

The dominant correction may thus be looked upon as interaction of a particle of a given bulk friction matrix $\bzz_0$ with an image particle via the propagator $\BG_w$, which accounts for the flow reflected by the wall. The matrix elements of ${\BG_w}$ decay according to their multipole indices as
\begin{equation}\label{matrixelementsG}
\BG_w^{ab}(\BR=2H\n) = \frac{1}{8\pi\eta}\left(\frac{1}{2H}\right)^{l+l'+\sigma+\sigma'-1}\bg^{ab}(\n),
\end{equation}
with the multipolar indices denoted by $a,b\in\{t,r,d\}$. The indices $l,\sigma$ refer to the superscript $a$, while $l'\sigma'$ refer to $b$. The directional tensors $\bg(\n)$ depend only on the direction of the normal vector $\n$. To derive the correction for an axisymmetric particle, we need the following elements
\begin{eqnarray}\label{gggg}
 g^{tt}_{\alpha\beta}  &=& - {3} n_\alpha n_\beta - \frac{3}{2}(\delta_{\alpha\beta}-n_\alpha n_\beta), \\ \nonumber
g^{rr}_{\alpha\beta}  &=& -n_\alpha n_\beta - \frac{5}{2}(\delta_{\alpha\beta}-n_\alpha n_\beta), \\ \nonumber
g^{dr}_{\alpha\beta\gamma} &=& - g^{rd}_{\gamma\beta\alpha} = 3 {\overbracket{n_\alpha\epsilon_{\beta\gamma\sigma}}}^{(\alpha\beta)} n_\sigma , \\ \nonumber
g^{dt}_{\alpha\beta\gamma} &=& g^{td}_{\gamma\alpha\beta} = \frac{9}{2}  {\overbracket{ n_\alpha n_\beta}}^{(\alpha\beta)} n_\gamma+  3   {\overbracket{n_\alpha(\delta_{\beta\gamma}- n_\beta n_\gamma)}}^{(\alpha\beta)}, \\ \nonumber
g^{dd}_{\alpha\beta\gamma\nu} &=& -\frac{3}{2} {\overbracket{ n_\alpha n_\beta}}^{(\alpha\beta)}{\overbracket{ n_\gamma n_\nu}}^{(\gamma\nu)} -12 {\overbracket{ \delta_{\alpha\gamma} n_\beta n_\nu}}^{(\alpha\beta)(\gamma\nu)} \\ \nonumber
&\phantom{=}&  -3  {\overbracket{ \delta_{\alpha\gamma} \delta_{\beta\nu}}}^{(\alpha\beta)(\gamma\nu)}.
\end{eqnarray}
The bar indicates the symmetric and traceless part of the respective tensors with respect to the indices in brackets.  The symbol ${\overbracket{\phantom{abc}}}^{(\alpha\beta)}$ indicates the symmetric and traceless part in the index pair $(\alpha,\beta)$. The appropriate reductions read 
\begin{align}
&{\overbracket{ n_\alpha n_\beta}}^{(\alpha\beta)} =  n_\alpha n_\beta - \tfrac{1}{3}\delta_{\alpha\beta},  \\ \nonumber
&{\overbracket{ \delta_{\alpha\gamma} \delta_{\beta\nu}}}^{(\alpha\beta)(\gamma\nu)} = \tfrac{1}{2}(\delta_{\alpha\gamma}{\delta_{\beta\nu}}+\delta_{\alpha\nu}\delta_{\beta\gamma}) - \tfrac{1}{3}\delta_{\alpha\beta}\delta_{\gamma\nu},  \\ \nonumber
&{\overbracket{ n_\alpha (\delta_{\beta\gamma}-n_\beta n_\gamma)}}^{(\alpha\beta)} = \tfrac{1}{2}(n_\alpha \delta_{\beta\gamma}+ n_\beta \delta_{\alpha\gamma}) - n_\alpha n_\beta n_\gamma, & \\ \nonumber
&{\overbracket{ n_\alpha \epsilon_{\beta\gamma\sigma}}}^{(\alpha\beta)} n_\sigma = \tfrac{1}{2}(n_\alpha \epsilon_{\beta\gamma\sigma} n_\sigma + n_\beta \epsilon_{\alpha\gamma\sigma} n_\sigma),  \\ \nonumber
&{\overbracket{ \delta_{\alpha\gamma} n_\beta n_\nu }}^{(\alpha\beta)(\gamma\nu)} = \tfrac{1}{9}\delta_{\alpha\beta}\delta_{\gamma\nu} - \tfrac{1}{3}(\delta_{\alpha\beta} n_\gamma n_\nu + n_\alpha n_\beta \delta_{\gamma\nu}) \\ \nonumber
& + \tfrac{1}{4} (\delta_{\alpha\gamma} n_\beta n_\nu + \delta_{\alpha\nu} n_\beta n_\gamma
 + \delta_{\beta\gamma} n_\alpha n_\nu + \delta_{\beta\nu} n_\alpha n_\gamma). 
\end{align}

Taking into account the symmetries of the bulk friction matrix $\bzz_0$ of an axisymmetric particle (i.e. the lack of $tr$ and $td$ elements), we find explicit expression for the correction terms in Eqs. \eqref{corrtt}-\eqref{corrrr} as
\begin{align}\label{tensors_abc} 
\BA_1 &= \ztt_0\bg^{tt}\ztt_0, \\ \nonumber
\BA_2 &= \bzz^{tt}_0\bg^{tt}\bzz^{tt}_0\bg^{tt}\bzz^{tt}_0, \\ \nonumber
\BB &= \ztt_0\bg^{td}\bzz^{dr}_0,\qquad \BB^\mathrm{T} = \bzz^{rd}_0\bg^{dt}\ztt_0, \\ \nonumber
\BC &= \zrr_0\bg^{rr}\zrr_0 + \zrr_0\bg^{rd}\bzz^{dr}_0 + \bzz^{rd}_0\bg^{dr}\zrr_0+\bzz^{rd}_0\bg^{dd}\bzz^{dr}_0, \label{tensors_end}
\end{align}
where appropriate contractions of the tensors are taken. The evaluation of these expressions using Eqs. \eqref{gggg} and the general form of $\bzz_0$ (cf. Ref. \onlinecite{KimKarrila}) leads to the expressions \eqref{angulartt}-\eqref{angularrr}.

%\bibliography{bibliography}

\begin{thebibliography}{10}

\bibitem{lang2015soft}
P.~Lang and Y.~Liu, editors,
\newblock {\em Soft Matter at Aqueous Interfaces}, volume 917 of {\em Lecture
  Notes in Physics},
\newblock Springer International Publishing, 2016.

\bibitem{HappelBrenner}
J.~Happel and H.~Brenner,
\newblock {\em Low Reynolds Numbers Hydrodynamics},
\newblock Kluwer, Dordrecht, 1991.

\bibitem{Banerjee2005}
A.~Banerjee and K.~Kihm,
\newblock Phys. Rev. E {\bf 72}, 042101 (2005).

\bibitem{Kihm2004}
K.~D. Kihm, A.~Banerjee, C.~K. Choi, and T.~Takagi,
\newblock Exp. Fluids {\bf 37}, 811 (2004).

\bibitem{Prieve1987}
D.~C. Prieve, F.~Lanni, and F.~Luo,
\newblock Faraday Discuss. Chem. Soc. {\bf 83}, 297 (1987).

\bibitem{Prieve1999}
D.~C. Prieve,
\newblock Adv. Coll. Interf. Sci. {\bf 82}, 93  (1999).

\bibitem{Walz1995}
J.~Y. Walz and L.~Suresh,
\newblock J. Chem. Phys. {\bf 103} (1995).

\bibitem{Lin2000}
B.~Lin, J.~Yu, and S.~Rice,
\newblock Phys. Rev. E {\bf 62}, 3909 (2000).

\bibitem{Lan1986}
K.~H. Lan, N.~Ostrowsky, and D.~Sornette,
\newblock Phys. Rev. Lett. {\bf 57}, 17 (1986).

\bibitem{Holmqvist2006}
P.~Holmqvist, J.~K.~G. Dhont, and P.~R. Lang,
\newblock Phys. Rev. E {\bf 74}, 021402 (2006).

\bibitem{Holmqvist2007}
P.~Holmqvist, J.~K.~G. Dhont, and P.~R. Lang,
\newblock J. Chem. Phys. {\bf 126}, 044707 (2007).

\bibitem{Lisicki2012}
M.~Lisicki, B.~Cichocki, J.~K.~G. Dhont, and P.~R. Lang,
\newblock J. Chem. Phys. {\bf 136}, 204704 (2012).

\bibitem{Rogers2012}
S.~A. Rogers, M.~Lisicki, B.~Cichocki, J.~K.~G. Dhont, and P.~R. Lang,
\newblock Phys. Rev. Lett. {\bf 109}, 098305 (2012).

\bibitem{Lisicki2014}
M.~Lisicki, B.~Cichocki, S.~Rogers, J.~K.~G. Dhont, and P.~R. Lang,
\newblock Soft Matter {\bf 10}, 4312 (2014).

\bibitem{Haghighi2013}
M.~Haghighi, M.~N. Tahir, W.~Tremel, H.-J. Butt, and W.~Steffen,
\newblock J. Chem. Phys. {\bf 139}, 064710 (2013).

\bibitem{Russel1977}
W.~B. Russel, E.~J. Hinch, L.~G. Leal, and G.~Tieffenbruck,
\newblock J. Fluid Mech. {\bf 83}, 273 (1977).

\bibitem{Mitchell2015}
W.~H. Mitchell and S.~E. Spagnolie,
\newblock J. Fluid. Mech. {\bf 772}, 600 (2015).

\bibitem{Katz1975}
D.~F. Katz, J.~R. Blake, and S.~L. Paveri-Fontana,
\newblock J. Fluid Mech {\bf 72}, 529 (1975).

\bibitem{Yang1983}
S.~M. Yang and L.~G. Leal,
\newblock J. Fluid Mech. {\bf 136}, 393 (1983).

\bibitem{DeMestre1975}
N.~J. {De Mestre} and W.~B. Russel,
\newblock J. Eng. Math. {\bf 9}, 81 (1975).

\bibitem{Hsu1989}
R.~Hsu and P.~Ganatos,
\newblock J. Fluid Mech. {\bf 207}, 29 (1989).

\bibitem{DeCorato2015}
M.~{De Corato}, F.~Greco, G.~D’Avino, and P.~L. Maffettone,
\newblock J. Chem. Phys. {\bf 142}, 194901 (2015).

\bibitem{Padding2010}
J.~T. Padding and W.~J. Briels,
\newblock J. Chem. Phys. {\bf 132}, 054511 (2010).

\bibitem{Cichocki2000}
B.~Cichocki, R.~B. Jones, R.~Kutteh, and E.~Wajnryb,
\newblock J. Chem. Phys. {\bf 112}, 2548 (2000).

\bibitem{KimKarrila}
S.~Kim and S.~J. Karrila,
\newblock {\em Microhydrodynamics: Principles and Selected Applications},
\newblock Butterworth-Heinemann, Boston, 1991.

\bibitem{EkielJezewska2009}
M.~L. Ekiel-Je\.{z}ewska and E.~Wajnryb,
\newblock {Precise Multipole Method for Calculating Hydrodynamic Interactions
  Between Spherical Particles in the Stokes Flow},
\newblock in {\em Theoretical Methods for Micro Scale Viscous Flows}, edited by
  F.~Feuillebois and A.~Sellier, pages 127--172, 2009.

\bibitem{Lorentz1907}
H.~A. Lorentz,
\newblock {\em Abhandlung \"uber Theoretische Physik},
\newblock B. G. Teubner, Leipzig und Berlin, 1907.

\bibitem{Blake1971}
J.~R. Blake,
\newblock Proc. Camb. Phil. Soc. {\bf 70}, 303 (1971).

\bibitem{Cichocki1998}
B.~Cichocki and R.~B. Jones,
\newblock Physica A {\bf 258}, 273 (1998).

\bibitem{Cichocki1988pch}
B.~Cichocki, B.~U. Felderhof, and R.~Schmitz,
\newblock Physicochem. Hydrodyn. {\bf 10}, 383 (1988).

\bibitem{Cichocki1994g}
B.~Cichocki, B.~U. Felderhof, K.~Hinsen, E.~Wajnryb, and J.~Bławzdziewicz,
\newblock J. Chem. Phys. {\bf 100}, 3780 (1994).

\bibitem{Ortega2003}
A.~Ortega and J.~{Garc{\'{\i}}a de la Torre},
\newblock J. Chem. Phys. {\bf 119}, 9914 (2003).

\bibitem{Bedeaux1974}
D.~Bedeaux and P.~Mazur,
\newblock Physica A {\bf 78}, 505 (1974).

\bibitem{JonesSchmitz1988}
R.~Jones and R.~Schmitz,
\newblock Physica A {\bf 149}, 373 (1988).

\bibitem{Schmitz1978}
R.~Schmitz and B.~U. Felderhof,
\newblock Physica A {\bf 92}, 423 (1978).

\bibitem{Schmitz1982b}
R.~Schmitz and B.~U. Felderhof,
\newblock Physica A {\bf 113}, 90 (1982).

\bibitem{Schmitz1982d}
R.~Schmitz and B.~U. Felderhof,
\newblock Physica A {\bf 113}, 103 (1982).

\bibitem{EkielJezewska2011}
M.~L. Ekiel-Je\.{z}ewska and E.~Wajnryb,
\newblock Phys. Rev. E {\bf 83}, 067301 (2011).

\bibitem{Cichocki1999lub}
B.~Cichocki, M.~L. Ekiel-Jeżewska, and E.~Wajnryb,
\newblock J. Chem. Phys. {\bf 111}, 3265 (1999).

\bibitem{Durlofsky1987}
L.~Durlofsky, J.~F. Brady, and G.~Bossis,
\newblock J. Fluid Mech. {\bf 180}, 21 (1987).

\bibitem{Brady1988}
J.~F. Brady and G.~Bossis,
\newblock Annu. Rev. Fluid Mech. {\bf 20}, 111 (1988).

\bibitem{Felderhof1988}
B.~U. Felderhof,
\newblock Physica A {\bf 151}, 1 (1988).

\end{thebibliography}
%\bibliographystyle{phaip}

\end{document}